\documentclass[a4paper,12pt]{article}

\usepackage{amssymb}
\usepackage{amsfonts}
\usepackage{graphicx}

\usepackage[T2A]{fontenc}
\usepackage[cp1251]{inputenc}
\usepackage[bulgarian,english]{babel}

\usepackage[colorlinks=true, urlcolor=blue,  linkcolor=black, citecolor=black]{hyperref}
\usepackage{color}

\newtheorem{theorem}{Theorem}

\newtheorem{example}{Example}
\newtheorem{definition}{Definition}

\title{On some representations of context-free languages}
\author{Krasimir Yordzhev}

\date{}

\begin{document}
\maketitle

\begin{abstract}
Context-free languages are widely used to describe the syntax of programming languages and natural languages. Usually, we describe a context-free language mathematically with the help of context-free grammar (for generation) or pushdown automata (for recognition). The purpose of this study is to describe some unconventional methods of description of context-free languages, namely a representation with the help of finite digraphs and with automata – generators of context-free languages. We will mainly focus on the mathematical models of these representations.
\end{abstract}



\section{Introduction}

Let $\Sigma$ be a finite and non-empty set, which we will call {\it alphabet}. The elements of this set we will call {\it letters}.

We will call a \emph{word over the alphabet} $\Sigma$ each finite string of letters from $\Sigma$.
 A word that does not contain any letter is called an \emph{empty word}, which we will mark with $\varepsilon$. $\Sigma^*$ denotes the set of all words over $\Sigma$, including empty set. $\Sigma^+ =\Sigma^* \setminus \{ \varepsilon \}$. The term \emph{ length of a word} refers to the number of letters in it. The length of the word $\alpha$ will be expressed by $|\alpha |$. By definition $|\varepsilon |=0$.

Let $\alpha$ and $\beta$  be two words over the Alphabet $\Sigma$. Then $\alpha\beta$ denotes the \emph{ concatenation} of $\alpha$ and $\beta$, that is, the word formed by making a copy of $\alpha$ and following it by a copy of $\beta$. So $\Sigma^*$ is the free monoid with identity $\varepsilon$.

Let $\Sigma$ be an alphabet. Each subset  $L$ of $\Sigma^*$ is called \emph{ formal language} (or only \emph{ language})  over alphabet $\Sigma$. A language $L$ is $\varepsilon$-\emph{free} if $\varepsilon \notin L$.

A \emph{context-free grammar} $\Gamma$ is the triple $\Gamma = \langle \mathcal{N} ,\Sigma ,\mathcal{P} \rangle$, where $\mathcal{N}$, $\Sigma$  are finite sets of nonterminals and terminals, respectively, $\mathcal{N} \cap \Sigma =\emptyset$ and $\mathcal{P}$ is a finite subset of the Cartesian product $\mathcal{N}\times \left( \mathcal{N} \cup \Sigma \right)^*$, whose elements are called \emph{productions} or \emph{rules}. The elements of $\mathcal{P}$ are denoted $A\to\omega$, where $A\in \mathcal{N}$. The notation $A\Rightarrow \omega$ indicates that there exists a sequence $A\to \alpha_1 A_1 \beta_1$, $A_1 \to \alpha_2 A_2 \beta_2$, $\ldots$ , $A_{t-2} \to \alpha_{t-1} A_{t-1} \beta_{t-1}$, $A_{t-1} \to \gamma$, where $A_i \in \mathcal{N}$ and $\alpha_i , \beta_i \in \left( \mathcal{N} \cup \Sigma \right)^*$ for every $i=1,2,\ldots t-1$, $\gamma \in \left( \mathcal{N} \cup \Sigma \right)^*$ and $\omega = \alpha_1 \alpha_1 \cdots \alpha_{t-1} \gamma \beta_{t-1} \beta_{t-2} \cdots \beta_1$. This sequence is called a \emph{derivation} with \emph{length} t.

Let $\Gamma = \langle \mathcal{N} ,\Sigma ,\mathcal{P} \rangle$ be a context-free  grammar an let $S\in \mathcal{N}$. Then the set $$L(\Gamma, S) = \left\{ \alpha \in \Sigma^* \; \; S\Rightarrow \alpha \right\}$$ is the language generated by the grammar $\Gamma$ with the \emph{starting symbol} $S$.

A context-free grammar $\Gamma = \langle \mathcal{N} ,\Sigma ,\mathcal{P} \rangle$ is $\varepsilon$-\emph{free} if $\varepsilon\notin L(\Gamma, S)$  for every $S\in \mathcal{N}$

It is well known  \cite{dshtr,Rayward-Smith,syt} that any context-free language can be generated by some grammar in \emph{Chomsky normal form}, i.e. a grammar in which all the productions have the form $A\to BC$ or $A\to a$, where $A,B,C\in\mathcal{N}$ are nonterminals and $a\in\Sigma$ is a terminal.

The algebraic properties of context-free grammars and languages are discussed in \cite{Chiswell,ginzburg,hopcroft,Lallement}. Several applications of formal grammars and languages and pushdown automata are considered in \cite{aho_ulman,hopcroft}.

A \emph{directed graph} or \emph{digraph} for short $G$ is a pair $G = \langle V,E\rangle$ where $V$ is a nonempty set, and $E$ is a multiset of ordered pairs of elements from $V$. The elements of $V$ are the \emph{vertices}  of the digraph G, the elements of E are its \emph{arcs} (or \emph{oriented edges}). A \emph{walk} of \emph{length} $t$ in a digraph $G$ is a non-empty alternating sequence $A_0 \rho_0 A_1 \rho_1 \ldots \rho_{t-1} A_t$ of vertices $A_i$, $i=0,1,\ldots ,t$ and arcs $\rho_i$, $i=0,1,\ldots ,t-1$ in $G$ such that $\rho_i =\langle A_i ,A_{i+1} \rangle$ for all $i < t$.

For more details on graph theory see \cite{diestel,harary} for example.

The widespread use of graph theory in different areas of science and technology is well known. For example, graph theory is a good tool for the modelling of computing devices and computational processes and in some non-traditional areas, such as social science or modelling some processes in education and other humanitarian activities \cite{harary1953graph,OrozKrAt,OrAtTo}. So many of graph algorithms have been developed \cite{swami}.

\emph{Transition diagram} is called a finite directed graph, all of whose arcs are labeled by an element of a semigroup. If $\pi$ is a walk in a transition diagram, then the \emph{label} of this walk $l(\pi)$ is the product of the labels of the arcs that make up this walk, taken in passing the arcs.
If $\pi_1$ and $\pi_2$ are arcs in a transition diagram, such that the end of $\pi_1$ coincides with the begin of $\pi_2$ and $\pi = \pi_1 \pi_2$, then $l(\pi )=l(\pi_1) l(\pi_2 )$.

A classic example of the representation of context-free languages using finite graphs is the transition diagram of pushdown automaton - recognizer of the corresponding context-free language. The paper \cite{RN77} describes a qualitatively new recognizer of context-free languages, based on some operations from graph theory. In the present article, we continue the work started in the mentioned above paper by improving the model and making it more user-friendly by adding new features and new useful tools.

Let $M$ be an arbitrary set. Throughout this paper $\mathcal{P} (M)$ will be the set of all subsets of $M$, including the empty one.

\section{A graph representation of context-free languages}

Let $\Sigma$ and $\cal N$ be finite sets and let $\Sigma^*$ be the free monoid over $\Sigma$ with identity the empty word $\varepsilon$. We define the set

\begin{equation}\label{N'}
  \mathcal{N}' = \left\{ A'\; |\; A\in \cal N \right\} ,\quad \cal N'\cap N =\emptyset .
\end{equation}

We define the monoid $T$ with the set of generators $\cal N\cup N'$ and the set of defining relations

\begin{equation}\label{AA'}
  AA'=\varepsilon ,
\end{equation}
where $A\in \mathcal{N}$, $A'\in \mathcal{N}'$ and $\varepsilon $ is the empty word (the identity of monoid $T$).

Let
\begin{equation}\label{W}
  W=\Sigma \times T=\left\{ \langle \alpha ,\omega \rangle \; |\; \alpha\in \Sigma^* ,\; \omega\in T\right\} .
\end{equation}

In $ B $ we define the operation
\begin{equation}\label{circ}
  \langle \alpha_1 ,\omega_1 \rangle \circ \langle \alpha_2 ,\omega_2 \rangle =\langle \alpha_1 \alpha_2 ,\omega_1 \omega_2 \rangle ,
\end{equation}
where $\alpha_1 ,\alpha_2 \in \Sigma^*$, $\omega_1 ,\omega_2 \in T$. It is easy to see that $W$ with so entered operation is monoid with unity element  $\langle \varepsilon ,\varepsilon \rangle$.

\begin{definition}\label{G_Gamma} \rm
Let $\Gamma =\langle \mathcal N, \Sigma ,\mathcal P \rangle$ be a grammar in Chomsky normal form. We construct the \emph{transition diagram}

\begin{equation}\label{GGamma}
  G_\Gamma =\langle V, E\rangle
\end{equation}
with the \emph{set of vertices}
\begin{equation}\label{V}
  V=\mathcal{N} \cup \{ Z\},\quad Z\notin \mathcal{N}
\end{equation}
and the \emph{multiset of arcs}
\begin{equation}\label{E}
  E\subseteq \left\{ \overrightarrow{AB}\; |\; A,B\in V \right\} .
\end{equation}
We \emph{label} the arcs of $G_\Gamma$ using the function
\begin{equation}\label{labeling}
  l : E \to \left\{ \langle a, \varepsilon \rangle \; | \; a\in \Sigma  \right\} \cup \left\{ \langle \varepsilon , X\rangle \; | \; X\in \mathcal{N}\cup \mathcal{N}' \right\} .
\end{equation}
Each arc in $G_\Gamma$ satisfies one of the following two conditions:
\begin{description}
  \item[(a)]\label{usl1} When $A\in \mathcal{N}$, $a\in\Sigma$, then there is an arc $\overrightarrow{AZ} \in E$ labeled $l(\overrightarrow{AZ}) = \langle a,\varepsilon\rangle$ if and only if $A\to a$ is a production in $\Gamma$;
  \item[(b)]\label{usl2} When $A,B,C\in \mathcal{N}$, then there are arcs $\overrightarrow{AB} \in E$ and $\overrightarrow{ZC} \in E$ with labels respectively $l(\overrightarrow{AB}) = \langle \varepsilon ,C\rangle$ and $l(\overrightarrow{ZC}) = \langle \varepsilon ,C' \rangle$ if and only if $A\to BC$ is a production in $\Gamma$.
\end{description}
\end{definition}

\begin{theorem}\label{th1}
Let $\Gamma =\langle \mathcal N, \Sigma ,\mathcal P \rangle$ be an $\varepsilon$-free grammar in Chomsky normal form and let $G_\Gamma$ be the transition diagram according to Definition \ref{G_Gamma}. Let $A\in \mathcal{N}$, $\alpha \in \Sigma^+$. Then $\alpha\in L(\Gamma , A)$ if an only if there is a walk $\pi$ with begin vertex $A$, end vertex $Z$ ($Z\notin \mathcal{N}$) and having label $l(\pi )=\langle \alpha , \varepsilon \rangle$.
\end{theorem}

Proof.
Necessity. Let $A\in \mathcal{N}$ and let $\alpha \in L(\Gamma ,A)$. Then there is a derivation $A\Rightarrow \alpha$. Let the length of this derivation be equal to $t\ge 1$. We will prove the necessity by induction on $t$.

Let $t=1$. Since $\Gamma$ is a grammar in Chomsky normal form, then $\alpha =a$, where $a\in \Sigma$, and $A\to a$ is a production from $\Gamma$. According to condition (a) in Definition \ref{G_Gamma}, in $G_\Gamma$ there is an arc $\overrightarrow{AZ}$ with label $l(\overrightarrow{AZ} )=\langle a,\varepsilon \rangle =\langle \alpha ,\varepsilon \rangle$. Therefore when $t = 1$ the necessity is fulfilled.

Suppose that for all $A\in N$ and for all $\alpha \in L(\Gamma, A)$ for which there is a derivation $A\Rightarrow \alpha$ with length not greater than $t$, in $G_\Gamma$ there is a walk  with the start vertex $A$, the final vertex $Z$ and having label $\langle \alpha , \varepsilon \rangle$.

Let $A\Rightarrow \alpha$ is a derivation in $\Gamma$  which length is equal to $t+1$ and let $A\to BC$, $A,B,C\in \mathcal{N}$ be the first production in this derivation. Then in $\Gamma$ there exist derivations $B\Rightarrow \alpha_1$ and $C\Rightarrow \alpha_2$ with lengths not greater than $t$, where $\alpha_1 ,\alpha_2 \in \Sigma^+$ and $\alpha_1 \alpha_2 =\alpha$. By the inductive assumption, in $G_\Gamma$ there are:

i) a walk $\pi_1$ with the start vertex $B$, final vertex $Z$, labeled $l(\pi_1 )=\langle \alpha_1 ,\varepsilon \rangle$ and

ii) a walk $\pi_2$ with start vertex $C$, final vertex $Z$ and labeled $l(\pi_2 )=\langle \alpha_2 ,\varepsilon \rangle$.

According to Definition \ref{G_Gamma}, condition (b), in $G_\Gamma$ there are arcs $\overrightarrow{AB}$ and $\overrightarrow{ZC}$ with labels $l(\overrightarrow{AB} )=\langle \varepsilon ,C\rangle$ and $l (\overrightarrow{ZC} )=\langle \varepsilon , C' \rangle$ respectively. Then the walk $\pi =\overrightarrow{AB}\pi_1 \overrightarrow{ZC}\pi_2$ has  start vertex $A$, final vertex $Z$ and label:

$l(\pi )=l(\overrightarrow{AB} )\circ l(\pi_1 )\circ l(\overrightarrow{ZC})\circ l(\pi_2 )=
\langle \varepsilon ,C\rangle \circ \langle \alpha_1 ,\varepsilon \rangle\circ \langle \varepsilon ,C' \rangle\circ \langle \alpha_2 ,\varepsilon \rangle =
\langle \varepsilon\alpha_1 \varepsilon\alpha_2 , C \varepsilon C' \varepsilon \rangle =\langle \alpha_1 \alpha_2 ,CC' \rangle =\langle \alpha ,\varepsilon \rangle .$

This proves the necessity.

Sufficiency. Let $A\in\mathcal{N}$ and let in $G_\Gamma$ there exists a walk $\pi$ with start vertex $A\in \mathcal{N}$, final vertex $Z\notin \mathcal{N}$ an label  $l(\pi)=\langle \alpha ,\varepsilon \rangle$, where $\alpha\in\Sigma^+ $. We will prove the sufficiency by induction on the length $|\alpha |$ of the word $\alpha$.

If $|\alpha |=1$, then $\alpha =a$ for some $a\in \Sigma$. Hence $\pi = \overrightarrow{AZ}$ (see Definition \ref{G_Gamma}), where $A\in \mathcal{N}$ and $\pi$ is an arc with label $l(\pi )=\langle a,\varepsilon \rangle$.  Then according to Definition \ref{G_Gamma}, condition (a), in $\Gamma$ there is a production $A\to a$, i.e. $\alpha=a\in L(\Gamma , A)$.

Let $t$ is a positive integer, such that for every vertex $A\in \mathcal{N}$ and every walk $\pi$ in $G_\Gamma$ with start vertex $A$,  final vertex $Z$ and label $l(\pi )=\langle \alpha ,\varepsilon \rangle$, $\alpha \in \Sigma^+$, from $|\alpha |\le t$ follows $\alpha\in L(\Gamma , A)$.

\begin{figure}[h]
\begin{center}
\includegraphics{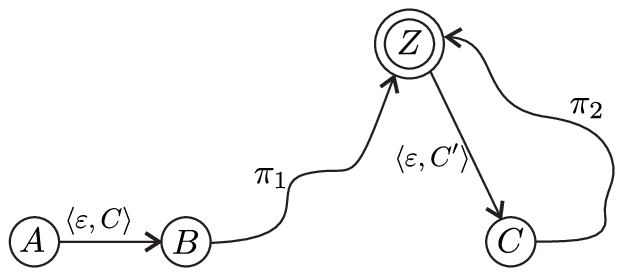}
\caption{}\label{fg1}
\end{center}
\end{figure}

Let $\alpha\in\Sigma^+$, where $|\alpha | = t+1\ge2$ and let $\pi$ be a walk in $G_\Gamma$ with start vertex $A\in \mathcal{N}$, final vertex $Z$ and label $l(\pi )=\langle \alpha ,\varepsilon \rangle$. Since $|\alpha |\ge 2$, there exists a vertex $B\in\mathcal{N}$ (i.e. $B\neq Z$), such that the first arc of $\pi$ is $\overrightarrow{AB}$ and let $l(\overrightarrow{AB} )=\langle \varepsilon ,C\rangle$, where $C\in\mathcal{N}$. But $l(\pi )=\langle \alpha ,\varepsilon  \rangle$. Therefore, in order for the letter $C$ to disappear from the label of $\pi$, it follows that in $G_\Gamma$ there exist an arc $\overrightarrow{ZC}$ with the label $\langle \varepsilon ,C' \rangle$ and walks  $\pi_1$ and $\pi_2$, where $\pi_1$ has  start vertex $B$, final vertex $Z$ and $\pi_2$ has start vertex $C$, final vertex $Z$, such that the path $\pi$ is represented in the form $\pi = \overrightarrow{AB} \pi_1 \overrightarrow{ZC} \pi_2$ (see Figure \ref{fg1}).

Let $l(\pi_1 )=\langle \alpha_1 ,\omega_1 \rangle$, $l(\pi_2 )=\langle \alpha_2 ,\omega_2 \rangle$, where $\alpha_1 ,\alpha_2 \in \Sigma^+$ and $\omega_1 ,\omega_1 \in \mathcal{N}^*$. Then
$$l(\pi )=l(\overrightarrow{AB} \pi_1 \overrightarrow{ZC} \pi_2 )= l(\overrightarrow{AB} )\circ l( \pi_1 )\circ l(\overrightarrow{ZC} )\circ l(\pi_2 )=$$
$$=\langle \varepsilon ,C\rangle \circ \langle \alpha_1 ,\omega_1 \rangle \circ \langle \varepsilon ,C' \rangle \circ \langle \alpha_2 ,\omega_2 \rangle = $$
$$=\langle \alpha_1 \alpha_2 ,C\omega_1 C' \omega_2 \rangle .$$

Since $l(\pi )=\langle \alpha ,\varepsilon \rangle$, we obtain $\alpha_1 \alpha_2 =\alpha$ and $C\omega_1 C'\omega_2 =\varepsilon$. It is easy to prove that the last equality is true if and only if $\omega_1 =\omega_2 =\varepsilon$. Since $|\alpha_1 |\ge 1$, $|\alpha_2 |\ge 1$ and $|\alpha_1 |+|\alpha_2 |=|\alpha |$, we have $|\alpha_1 |\le |\alpha |=t+1$ and $|\alpha_2 |\le |\alpha |=t+1$, i.e. $|\alpha_1 |\le t$ and $\alpha_2 \le t$. By the inductive assumption, $\alpha_1 \in L(\Gamma ,B)$ and $\alpha_2 \in L(\Gamma ,C$, i.e. in $\Gamma$ there exist derivations $B \Rightarrow \alpha_1$ and $C \Rightarrow  \alpha_2$. Therefore in $\Gamma$ there is a derivation $A\to BC \Rightarrow \alpha_1 C  \Rightarrow \alpha_1 \alpha_2 =\alpha$. This proves the sufficiency.

\hfill $\Box$

\begin{example}
Consider the context-free grammar in Chomsky normal form $\Gamma =\langle \{ S,A,B,C\}, \{ 0,1\} ,\{ S\to AB, B\to SC, S\to AC, A\to 0, C\to 1\} \rangle$. It is easy to prove that $L(\Gamma , S) =\left\{ 0^n 1^n \; |\; n=1,2,, \ldots \right\}$. This language is a classic example of a language that is context-free but not regular \cite{aho_ulman,dshtr,syt}. The corresponding graph $G_\Gamma$ is shown in Figure \ref{fg2}.

\begin{figure}[h]
\begin{center}
\includegraphics{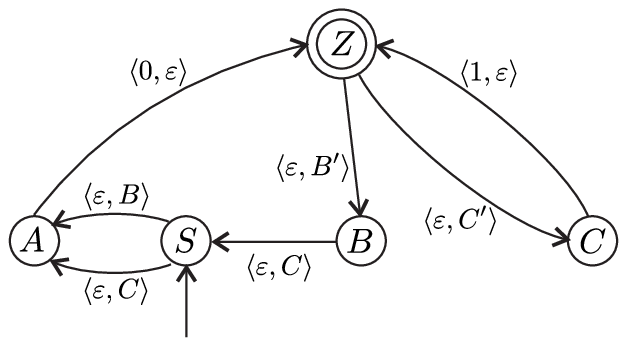}
\caption{}\label{fg2}
\end{center}
\end{figure}
\end{example}

\section{Nondeterministic finite automata -- generators of context-free languages}

The following theorem is in a sense the opposite of Theorem \ref{th1}:

\begin{theorem} \label{th2}
Let $\mathcal{N}$ и $\Sigma$ be finite nonempty sets, $\mathcal{N} \cap \Sigma =\emptyset$, $Z\notin \mathcal{N} \cup \Sigma$ and let $W$ be the monoid defined by equations (\ref{N'}) -- (\ref{circ}).We construct the transition diagram $H=\langle V,E,l\rangle$ with a set of vertices $V=\mathcal{N} \cup \{ Z\}$, a multiset of arcs $E$ whose elements are labeled with elements of the monoid $W$ and satisfying the following conditions:

a) If $A\in \mathcal{N}$ and $\overrightarrow{AZ}$ is an arc in $H$ then $l(\overrightarrow{AZ} )=\langle a, \varepsilon \rangle$ for some $a\in\Sigma$;

b) Let $A,B,C\in \mathcal{N}$. Then there exists an arc $\overrightarrow{AB}$ with label $l(\overrightarrow{AB} )=\langle \varepsilon , C\rangle $ if and only if there exists an arc $\overrightarrow{ZC}$ with label $l(\overrightarrow{ZC} )=\langle \varepsilon ,C' \rangle$;

c) $H$ contains no other arcs than those described in items a) and b).

Let $S\in\mathcal{N}$. By $\Pi_S$ denote the set of all walks in $H$  with start vertex $S$,  final vertex $Z$ and having a label of the type $\langle \alpha ,\varepsilon \rangle$, $\alpha\in\Sigma^+$. Then the language
$$L(H,S) =\left\{ \alpha \; | \; \exists \pi\in\Pi_S : l(\pi )=\langle \alpha ,\varepsilon \rangle \right\}$$
generated by the transition diagram $H$ with the starting vertex $S$ is context-free.
\end{theorem}

Proof. On the basis of the transition diagram $H$ we construct the context-free grammar $\Gamma =\langle \mathcal{N} , \Sigma ,\mathcal{P} \rangle$ in Chomsky normal form, where $\mathcal{P}$ consists of the productions:
\begin{itemize}
  \item $A\to a$, where $A\in\mathcal{N}$, $a\in\Sigma$ if and only if in $H$ there is an arc $\overrightarrow{AZ}$ with label $\langle a,\varepsilon \rangle$;
  \item $A\to BC$, where $A,B,C\in\mathcal{N}$ if and only if in $H$ there are an arc $\overrightarrow{AB}$ with label $\langle \varepsilon ,C\rangle$ and an arc $\overrightarrow{ZC}$ with label $\langle \varepsilon ,C' \rangle$.
\end{itemize}
It is easy to see that $\Gamma$ satisfies the conditions of Theorem \ref{th1}, has a transition diagram $G_\Gamma =H$ and $l(\Gamma ,S)=L(H,S)$ for every $S\in \mathcal{N}$.
 This completes the proof of Theorem \ref{th2}.

\hfill $\Box$

\begin{definition} \label{dfn2Y} \rm
\emph{Non-deterministic finite automaton generator of context-free languages} we will call the ordered 8-tuple
\begin{equation} \label{mathcalY}
\mathcal{Y} =\langle  \Sigma ,\mathcal{N} ,Z, V, S, W, \delta ,\lambda  \rangle ,
\end{equation}
where

$\Sigma$ is a finite set of \emph{terminal symbols};

$\mathcal{N}$ is a finite set of \emph{nonterminal symbols};

$Z\notin \mathcal{N}$ is a special symbol, which we will call \emph{final state};

$V=\mathcal{N} \cup \{ Z\}$ is a finite set of \emph{states};

$S\in \mathcal{N}$ is the \emph{start state};

$W$ is the monoid defined by the equations (\ref{N'}) -- (\ref{circ});

$\delta \; :\; V\to \mathcal{P} (V)$ is the \emph{transition function};

$\lambda \; :\; V\times \delta (V) \to \mathcal{P} (W)$ is the \emph{output (or labeling) function}, where $\delta (V)=\{ B\in V\; |\; \exists A\in V\; :\; B\in \delta (A)\}$, moreover $\lambda (A,B) $ is defined if and only if $B\in \delta (A)$.

The functions $\delta$ and $\lambda$  satisfy the following conditions:
\begin{enumerate}
  \item \label{df2i1} $Z\in \delta (A)$ if and only if $A\in \mathcal{N}$ and there exists $a\in\Sigma$ such that $\langle a,\varepsilon \rangle \in \lambda (A,Z)$;

  \item \label{df2i2} Let $A,B,C\in \mathcal{N}$ and let $B\in \delta (A)$. Then $\langle \varepsilon , C\rangle \in \lambda (A,B)$ if and only if $C\in \delta (Z)$ and $\langle \varepsilon , C' \rangle \in \lambda (Z,C)$;

  \item \label{df2i3} The functions $\delta$ and $\lambda $ are  not defined if they do not satisfy the condition \ref{df2i1} or the condition \ref{df2i2}.
\end{enumerate}
\end{definition}

\begin{definition}\label{sequen} \rm
Consider the sequences $A_0 ,A_1 ,\ldots , ,A_k$ and $w_1 ,w_2 ,\ldots w_k$, where
\begin{equation}\label{delta(Ai)}
  A_0 =S, A_1 \in\delta (A_0 ), \ldots ,A_i \in \delta (A_{i-1} ) ,\ldots ,A_k \in \delta (A_{k-1} )
\end{equation}
and
\begin{equation}\label{lambdaAiAi+1}
  w_1 \in \lambda (A_0 ,A_1 ), w_2 =\lambda (A_1 ,A_2 ),\dots  , w_k \in \lambda (A_{k-1} ,A_k )
\end{equation}

We say that the word  $\alpha \in\Sigma^+$  \emph{is generated by the automaton} (\ref{mathcalY}) if there exists a positive integer $k$ and sequences of the type (\ref{delta(Ai)}) and (\ref{lambdaAiAi+1}) such that
\begin{equation}\label{alphaepsilon}
w_k \in \langle \alpha ,\varepsilon \rangle .
\end{equation}
\end{definition}

\begin{definition} \label{ezikL} \rm
We say that the language  $L=L(\mathcal{Y} )$  \emph{is generated by the automaton} (\ref{mathcalY}) if $L$ consists of all words that are generated by the automaton (\ref{mathcalY}).
\end{definition}

It is easy to see that $\varepsilon \notin L(\mathcal{Y} )$, i.e.  the language $L(\mathcal{Y} )$ is $\varepsilon$-free for any automaton of the type (\ref{mathcalY}).

\begin{theorem}
Let $\mathcal{Y} =\langle  \Sigma ,\mathcal{N} ,Z, V, S, W, \delta ,\lambda  \rangle$ be an automaton of the type (\ref{mathcalY}). Then $L( \mathcal{Y} )$ is a context-free language.
\end{theorem}

Proof. We construct the transition diagram $G_\mathcal{Y} =\langle V,E,l\rangle$, where the set of vertices $ V $ coincides with the set of states of the automaton,  $\overrightarrow{AB} \in E$ and $l( \overrightarrow{AB} )=\langle \alpha ,\omega \rangle$ if and only if $B\in \delta (A)$ and $\langle \alpha ,\omega \rangle \in \lambda (A,B)$. It is easy to see that the transition diagram $G_\mathcal{Y}$ satisfies the conditions of Theorem \ref{th2}. This implies that $L(\mathcal{Y} )$ is a context-free language.

\hfill $\Box$


\end{document}